# Dirac points, new photonic band gaps and effect of magnetically induced transparency in dichroic cholesteric liquid crystals with wavelength dependent magnetooptical activity parameter

A.H. Gevorgyan

Far Eastern Federal University, Institute of High Technologies and Advanced Materials, 10 Ajax Bay, Russky Island, Vladivostok 690922, Russia

We investigated the properties of dichroic cholesteric liquid crystals (CLCs) being in external static magnetic field directed along helix axis. We have shown that in the case of the wavelength dependence of magneto-optic activity parameter, new features appear in the optics of dichroic CLCs. We have shown that in this case new Dirac points appear, moreover, at some Dirac points photonic band gaps (PBGs) appear, at others, lines of magnetically induced transparency (MIT). In this case a polarization-sensitive transmission band appears too. At certain values of the helix pitch of the CLC and of the magnitude of the external magnetic field three PBGs of different nature appear, a transmittance band, two narrow lines of MIT and one broadband MIT. This system is non-reciprocal and the nonreciprocity changes over a wide range, it is observed both for reflection and transmittance and for absorption. The soft matter nature of CLCs and their response to external influences lead to easily tunable multifunctional devices that can find a variety of applications. They can apply as tunable narrow-band or broad-band filters and mirrors, a highly tunable broad/narrow-band coherent perfect absorber, transmitter, ideal optical diode, and other devices.

## I. INTRODUCTION

Starting from pioneering papers by E. Yablonovich [1] and S. John [2], photonic crystals (PCs) are the objects of intensive theoretical and experimental research, because the results of such investigations find more and more applications in photonic devices of the new generation. Some PCs exhibit novel mechanisms of some classical quantum effects, such as Bloch oscillations (oscillations of electrons within the Brillouin zone induced by an applied electric field) [1], Zener tunneling (the direct tunneling of a Bloch particle into a continuum of another energy band, which takes place without extra energy in the presence of a large electric field in the crystal) [2], electromagnetically induced transparency (EIT; which is a quantum interference effect in three-level atomic systems that eliminates the absorption at the resonance frequency and gives rise to a narrow transparency window) [3,4], electromagnetically induced absorption (EIA; which is the counter phenomenon of invoking constructive interference between multiple interaction pathways to enhance/induce absorption) [5] and etc. In particular, some nanostructures in external magnetic field or nanostructures with strong magnetic dipole interactions exhibit a narrow transparency window, and some others a enhance/induce absorption. These new phenomena are called magnetically induced transparency (MIT), and magnetically induced absorption (MIA), respectively [6-15]. Magneto-optics of cholesteric liquid crystals (CLCs) is of particular interest. The papers [16-26] present the results of a theoretical and experimental study of the magneto-optical properties of CLCs. Then, in the papers [27-30] the existence of MIT and MIA was demonstrated in short wavelength part of spectrum in CLCs being in external magnetic field both at the presence and absence of local dielectric anisotropy. In [31] it was reported about observation of Dirac points in CLCs, both in the presence of external magnetic field and in its absence. It was shown that the MIT and MIA phenomena are observed exactly at the Dirac points.

However, all the works [16-31] deal with the case when the parameter of magneto-optical activity g is constant and is independent of wavelength. It was shown in [32] that in the case of the wavelength dependence of the magneto-optical activity parameter in the short-wavelength spectrum, no MIT effect nor MIA effect is observed. And now a natural question arises as to the existence of these effects in magnetically active CLCs in the case of the frequency dependence of the magneto-optical activity parameter. In this paper we show the existence of a new region of the photonic band gap (PBG) and the MIT effect in the case in the absence of local anisotropy in CLCs and present the results of a study of their features.

## II. MODELS AND METHODOLOGY. RESULTS

If the CLC is in external magnetic field directed along the helix axis and the medium has a magneto-optical activity, then the tensors of dielectric permittivity and magnetic permeability will have the forms:

$$\hat{\varepsilon}(z) = \varepsilon_m \begin{pmatrix} 1 + \delta\cos2az & \pm\delta\sin2az \pm ig/\varepsilon_m & 0 \\ \pm\delta\sin2az \mp ig/\varepsilon_m & 1 - \delta\cos2az & 0 \\ 0 & 0 & 1 - \delta \end{pmatrix}, \text{ and } \hat{\mu}(z) = \hat{I}, \quad (1)$$

where g is the parameter of magneto-optical activity of CLC, and it, in general, is a function of the external magnetic field, Verdet constant, and dielectric permittivity of media, $\varepsilon_m = (\varepsilon_1 + \varepsilon_2)/2$, $\delta = \frac{(\varepsilon_1 - \varepsilon_2)}{(\varepsilon_1 + \varepsilon_2)}$, $\varepsilon_{1,2}$ are the principal values of the local dielectric permittivity tensor in the presence of an external magnetic field, $a = 2\pi / p$, $p$ is the pitch of the helix, axis $z$ and external magnetic field are directed along CLC helix axis. The dependence of g on wavelength $\lambda$ follows directly from the following considerations. First, the magnitude of Faraday rotation angle for isotropic magnetoactive medium is given by the formula

$$\varphi = \frac{\pi d}{\lambda}\left(\sqrt{\varepsilon + g} - \sqrt{\varepsilon - g}\right), \quad (2)$$

where $d$ is the transmission distance and $\varepsilon$ is the dielectric permittivity. On the other hand, for the same magnitude of rotation experimenters use the formula $\varphi = VB_{ext}d$, where $V$ is the Verdet constant, and $B_{ext}$ is the external magnetic field induction. Substituting $\varphi = VB_{ext}d$ into (2) for g we will have

$$g = \frac{VB_{ext}\lambda}{\pi}\sqrt{\left|\varepsilon - \left(\frac{VB_{ext}\lambda}{2\pi}\right)^2\right|}. \quad (3)$$

The dispersion equation now has the form:

$$\left(\frac{\omega^2}{c^2}\varepsilon_1 - k_{mz}^2 - a^2\right)\left(\frac{\omega^2}{c^2}\varepsilon_2 - k_{mz}^2 - a^2\right) - \left(2ak_{mz} - \frac{\omega^2}{c^2}g\right)^2 = 0. \quad (4)$$

We will consider the limiting case, namely, the case of CLC in the absence of local birefringence, when $\text{Re}\Delta = \frac{\text{Re}\varepsilon_1 - \text{Re}\varepsilon_2}{2} = 0$, and $\text{Im}\Delta = \frac{\text{Im}\varepsilon_1 - \text{Im}\varepsilon_2}{2} \neq 0$, that is we consider the case $\varepsilon_{1,2} = \varepsilon + i\varepsilon_{1,2}''$, where $\varepsilon$ is the real part of the components of the dielectric constant tensor, which is assumed to be the same for all components. Consideration of this limiting case, below, will allow us to reveal many features of magnetooptical CLCs in their purest form, which in the more general case of $\text{Re}\Delta \neq 0$ have more complex manifestations. Some features of the optical properties of such media are considered in detail in [33-35]. In this case Eq. (4) splits into two quadratic equations that have roots, in the form:

$$k_{1z} = \frac{\omega}{c}\sqrt{\varepsilon - g} + a, \quad k_{2z} = \frac{\omega}{c}\sqrt{\varepsilon + g} - a, \quad (5)$$
$$k_{3z} = -\frac{\omega}{c}\sqrt{\varepsilon - g} + a, \quad k_{4z} = -\frac{\omega}{c}\sqrt{\varepsilon + g} - a.$$

Fig.1 shows the dependence of g on wavelength $\lambda$. The function $g(\lambda)$ has two zeros on the wavelengths $\lambda_1 = 0$ and $\lambda_2 = \frac{2\pi}{VB_{ext}}\sqrt{\varepsilon}$, and passes across local maximum on the wavelength $\lambda_3 = \frac{\pi}{VB_{ext}}\sqrt{2\varepsilon}$.

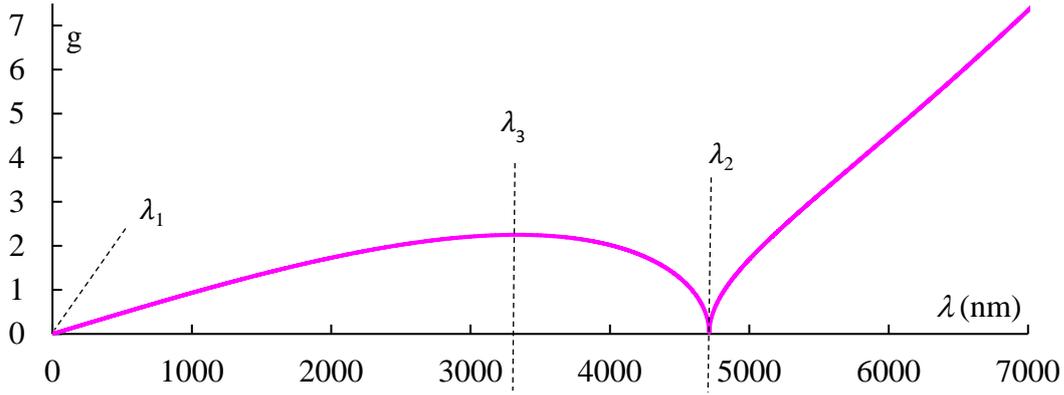

Fig.1. The dependence of g on wavelength $\lambda$. The parameters are: $\varepsilon = 2.25$; $V = \text{const} = 5 \cdot 10^5$ rad/(T · m); $B_{ext} = 40$ T.

Substituting $\lambda_3 = \frac{\pi}{VB_{ext}}\sqrt{2\varepsilon}$ into (3) for g at the wavelength $\lambda_3$ we will have g = $\varepsilon$. From (5) it follows that at this wavelength we will have $k_{1z} = k_{3z}$. Taking into account approximately linear dependence of $k_{iz}$ onto frequency $\omega$ near the wavelength $\lambda_3$, here at the wavelength $\lambda_3$ we have the Dirac point and here the touching of the vertices of the Dirac cones takes place. Now we pass to investigate the optical properties of dichroic CLCs near these points.

Fig. 2 shows the dependences of real and imaginary parts of wave vectors $k_{iz}$ on wavelength $\lambda$. As can be seen from these graphs, there is one more characteristic wavelength, namely $\lambda_4$, where the intersection of curves $\text{Re}k_{2z}(\lambda)$ and $\text{Re}k_{3z}(\lambda)$ occurs. Again, taking into account approximately linear dependence of $k_{iz}$ onto frequency $\omega$ near the wavelength $\lambda_4$, here at the wavelength $\lambda_4$ again we have the Dirac point. By equating $k_{2z}$ to $k_{3z}$ from (5) and taking into account (3) we obtain the expression for the wavelength for this Dirac point: $\lambda_4 = 2\pi p\sqrt{\frac{\varepsilon}{4\pi^2+V^2B_{ext}^2p^2}}$. At the presence of anisotropic absorption ($\text{Re}\varepsilon_1 = \text{Re}\varepsilon_2 = \varepsilon$, but $\text{Im}\varepsilon_1 \neq \text{Im}\varepsilon_2$) a PBG is formed here, which exists also at the absence of external magnetic field.

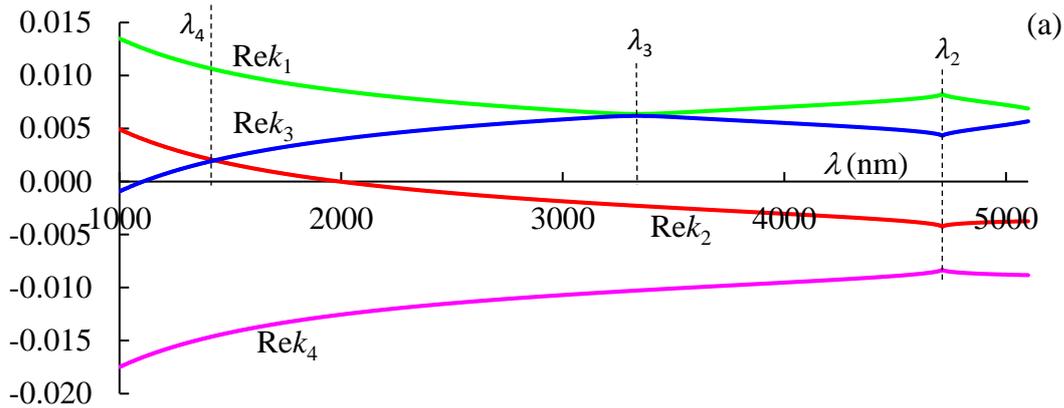

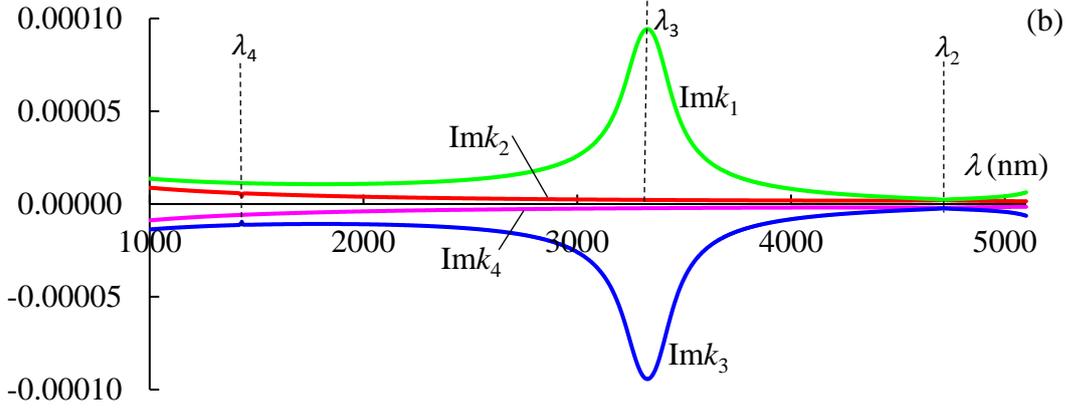

Fig. 2. The dependences of (a) real and (b) imaginary parts of wave vectors $k_{iz}$ on wavelength $\lambda$. $p = 1000$ nm, $\text{Im}\varepsilon_1 = 0.01$, $\text{Im}\varepsilon_2 = 0$. The other parameters are the same as in Fig.1.

Now, using the exact analytical solution of the Maxwell's equations for magnetoactive CLC [16], and dispersion equation (4) we can solve the problem of light reflection, transmission, and absorption in the case of a planar magnetoactive CLC layer of finite thickness. We assume that the optical axis of this CLC layer is perpendicular to the boundaries of the layer and is directed along the $z$-axis. The CLC layer on its both sides border with isotropic half-spaces with the same refractive indices equal to $n_s$. The boundary conditions, consisting of the continuity of the tangential components of the electric and magnetic fields, are a system of eight linear equations with eight unknowns (in more details see [16]). Solving this boundary-value problem, one can determine the values of the reflected $\mathbf{E}_r$ and transmitted $\mathbf{E}_t$ fields and calculate the energy coefficient of reflection $R = \frac{|\mathbf{E}_r|^2}{|\mathbf{E}_i|^2}$, transmission $T = \frac{|\mathbf{E}_t|^2}{|\mathbf{E}_i|^2}$, and absorption $A = 1 - (R + T)$, where $\mathbf{E}_i$ is the incident light field. Here and below, we will consider the case of minimal influence of dielectric boundaries, that is the case $n_s = \sqrt{\varepsilon_m}$.

Fig. 3 shows the spectra of reflection $R$ (curves 1 and 2), transmission $T$ (curves 3 and 4) and absorption $A$ (curves 5 and 6) at different directions of external magnetic field. The incident light has polarization coinciding with the first eigen polarization (EP) (curves 1,3,5) and with the second EP (curves 2, 4,6). In Fig. 3(a) the directions of the incident light and the external magnetic field coincide; and in Fig. 3(b) they are oppositely directed. By definition, the EPs are the two polarizations of incident light that do not change as the light passes through the system. These two EPs approximately coincide with orthogonal circular polarizations. Some discrepancy arises only near the wavelength $\lambda_2$. The helix of our CLC is right-handed. Let us now enumerate the EPs in the following way, we will assume that the first EP is the EP that approximately coincides with the right-hand circular polarization, while the second EP with the left-hand one.

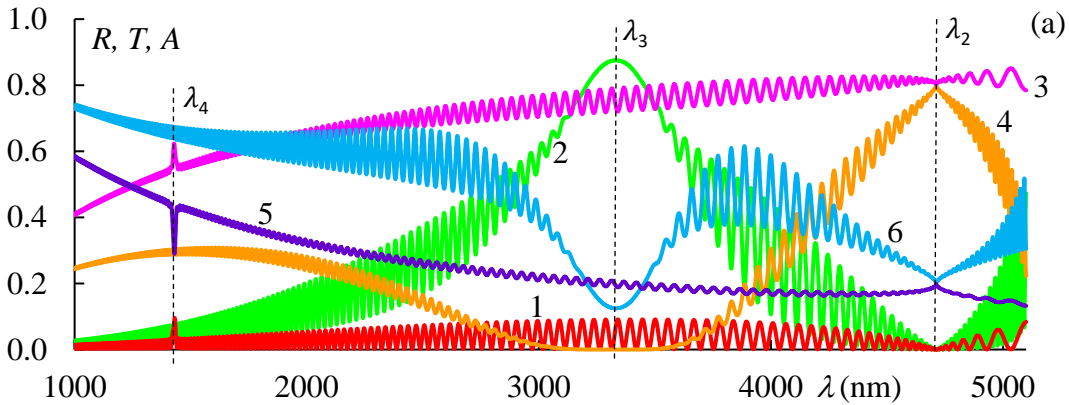

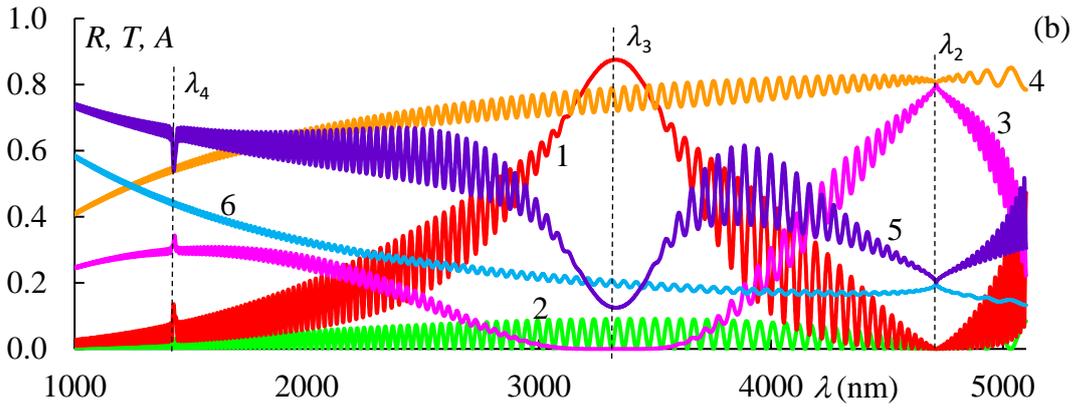

Fig.3. The spectra of reflection $R$ (curves 1 and 2), transmission $T$ (curves 3 and 4) and absorption $A$ (curves 5 and 6) at different directions of external magnetic field. The incident light has polarization coinciding with the first EP (curves 1,3,5) and with the second EP (curves 2, 4,6). (a): the directions of the incident light and the external magnetic field coincide; (b): they are oppositely directed. CLC layer thickness $d=50p$. The other parameters are the same as in Fig.2.

As can be seen from Fig. 3, at the wavelength $\lambda_4$, diffraction reflection undergoes light with a polarization coinciding with the first EP. This photonic band gap (PBG) exists also at the absence of external magnetic field and in this case $\lambda_4 = p\sqrt{\varepsilon}$. At the presence of local birefringence there arouse a PBG with a finite frequency width with $\Delta\lambda = p\Delta n = p(\sqrt{\mathrm{Re}\varepsilon_1} - \sqrt{\mathrm{Re}\varepsilon_2})$. Once again, let us note that the PBG arising in this case is determined by the structure of the CLC, i.e., the chirality sign of the polarization of the incident diffracting light is determined only by the chirality sign of the CLC helix. We call this PBG the first (or basic) PBG.

A new PBG is formed near the wavelength $\lambda_3$. We call this PBG the second PBG. This new PBG is sensitive to the polarization of the incident light too. But if the chirality sign of the polarization of the incident diffracting light for the first PBG is determined only by the chirality sign of the CLC helix, while for the second one it is determined by the external magnetic field direction (i.e., on whether the directions of the external magnetic field and the incident light are parallel, or they are antiparallel). In the first case, the diffraction reflection undergoes the light with the first EP, and in the second case, the light with the second EP.

Near the wavelength $\lambda_2$ a transmission band is formed for incident light with a certain polarization. And as for the second PBG the chirality sign of the polarization of the incident light for which transmission band were formed is determined by the external magnetic field direction.

Then, Fig. 4 shows the reflection, transmission and absorption spectra at different values of helix pitch. The incident light has polarization coinciding with the first EP ($R_1, T_1, A_1$) and with the second EP ($R_2, T_2, A_2$). In Fig. 4 (a, c, e, g, i, k) the directions of the incident light and the external magnetic field coincide, while in Fig. 1 (b, d, f, h, j, l) they are oppositely directed.

From the presented spectra, it follows that:
1) this system is non-reciprocal, in Fig. 4 the spectra in the left column differ significantly from those in the right column, the nonreciprocity changes over a wide range, it is observed both for reflection and transmittance and for absorption.
2) with an increase in the pitch of the helix, the height of the peaks of reflection of the first PBG decreases (in Fig. 4 a, b, these peaks are highlighted by brown arrows).
3) at the given parameters of the problem in the region $\lambda \geq \lambda_6 = 5180$ nm (about $\lambda_6$ see below) a new PBG (the third PBG) is formed, which is sensitive to the polarization of the incident light (see Fig. 4 b, c)
4) with an increase in the pitch of the helix, the height of the peaks of transmission of the first PBG on the wavelength $\lambda_4$ increases for the first EP (in Fig. 4 e these peaks are highlighted by brown arrows). Simultaneously, an increase in the height of the dips in the absorption on this wavelength for the first EP takes place, too (in Fig. 4 i these dips are highlighted by brown arrows) and as noted above in point 2), there is a decrease in the height of the reflection peaks on this wavelength. Thus, we have the following picture: with an increase in the pitch of the helix, the PBG at wavelength $\lambda_4$ gradually turns into a narrow window of MIT, i.e., at the Dirac point, for some parameters of the problem, diffraction reflection is observed, and for others, a narrow window of MIT with a continuous transition from one state to another.

5) with an increase in the pitch of the helix, the height of the peaks of transmission band near the wavelength $\lambda_2$ decreases (see Fig. 4 g and h).

6) at a certain value of the helix pitch, instead of a narrow line of magneto-induced transmission, a broadband region of magneto-induced transmission is formed (see Fig. 4 e and i, curve 3).

7) at certain values of the external magnetic field and helix pitch, a new MIT line appears in the spectra (it is indicated by a blue arrow in Figs. 4e and i). As our calculations show, there is a new Dirac point here. It appears due to the intersection of the curves $\mathrm{Re}k_{1z}(\lambda)$ and $\mathrm{Re}k_{2z}(\lambda)$. By equating $k_{1z}$ to $k_{2z}$ from (5) taking into account (3) we obtain the expression for the wavelength for this Dirac point: $\lambda_5 = 2\pi p\sqrt{\frac{\varepsilon(4\pi^2+V^2B_{ext}^2p^2)}{16\pi^4+V^4B_{ext}^4p^4}}$. Fig. 5 shows the dependences of real and imaginary parts of wave vectors $k_{iz}$ on wavelength $\lambda$ in the situation where there is an intersection the curves $\mathrm{Re}k_{1z}(\lambda)$ and $\mathrm{Re}k_{2z}(\lambda)$.

8) as follows from (5) at $g \geq \varepsilon$ two from four wave vectors become complex at the absence of absorption and a new (the third) PBG will appears. From condition $g = \varepsilon$ we obtain two wavelengths, namely $\lambda_3 = \frac{\pi}{VB_{ext}}\sqrt{2\varepsilon}$ and $\lambda_6 = \frac{\pi\sqrt{2\varepsilon}}{VB_{ext}}\sqrt{1+\sqrt{2}}$. As mentioned above, at the local maximum of $g(\lambda)$ we have $g = \varepsilon$, too.

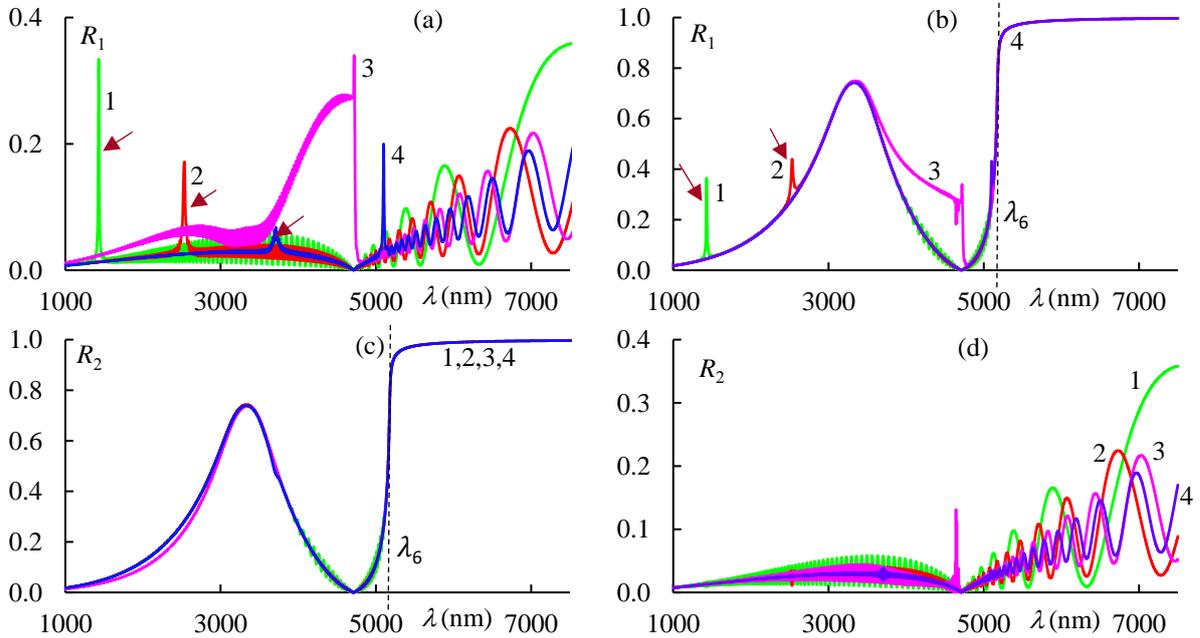

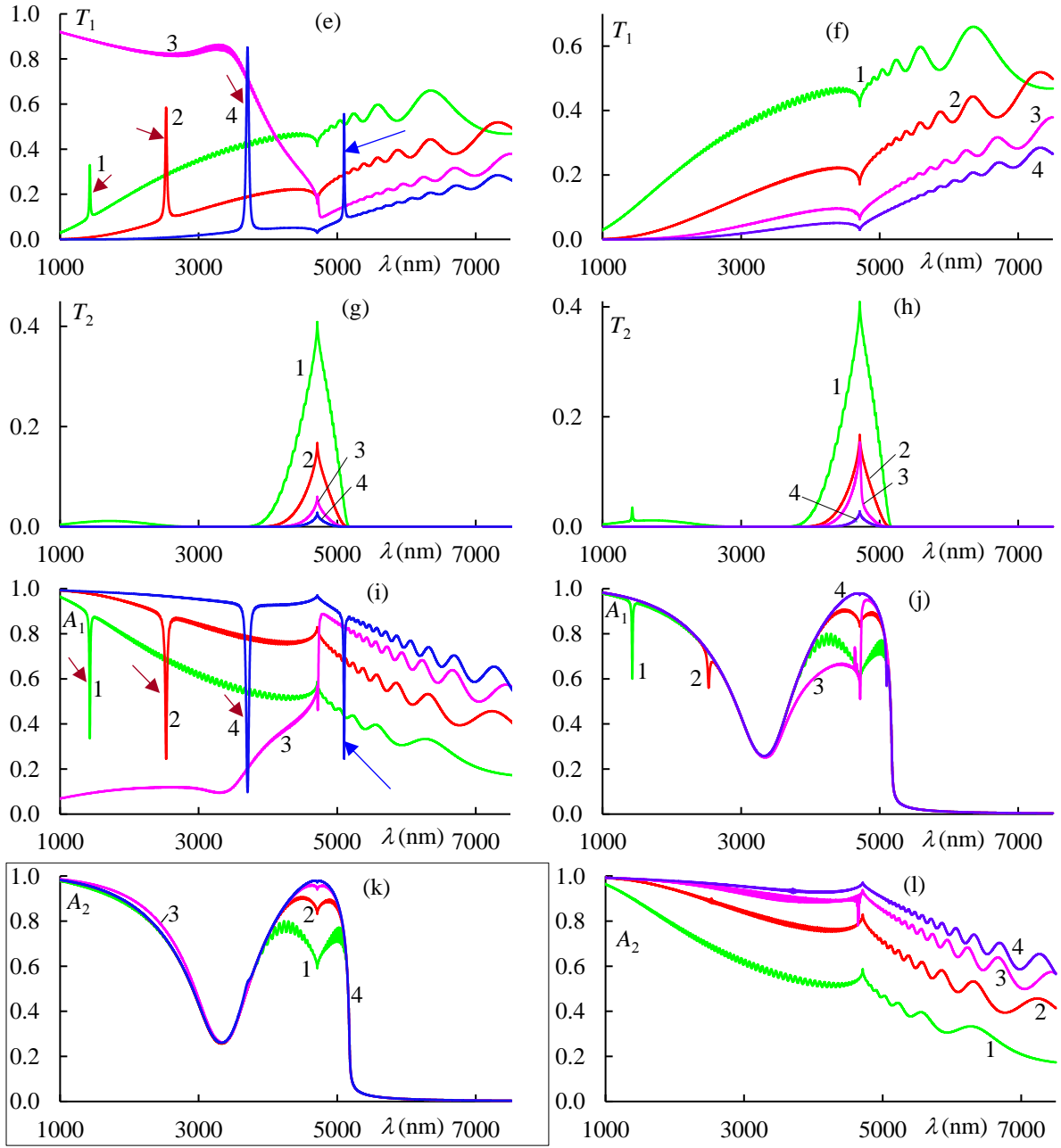

Fig. 4. The spectra of reflection, transmission and absorption at different values of helix pitch. The incident light has polarization coinciding with the first EP ($R_1, T_1, A_1$) and with the second EP ($R_2, T_2, A_2$). In Fig. 1 (a, c, e, g, i, k) the directions of the incident light and the external magnetic field coincide, while in Fig. 1 (b, d, f, h, j, l) they are oppositely directed. $p = 1000$ nm (curves 1), $p = 2000$ nm (curves 2), $p = 3150$ nm (curves 3) and finally $p = 4000$ nm (curves 4). $\text{Im}\varepsilon_1 = 0.05$, $\text{Im}\varepsilon_2 = 0$, $d=50p$. The other parameters are the same as in Fig.2.

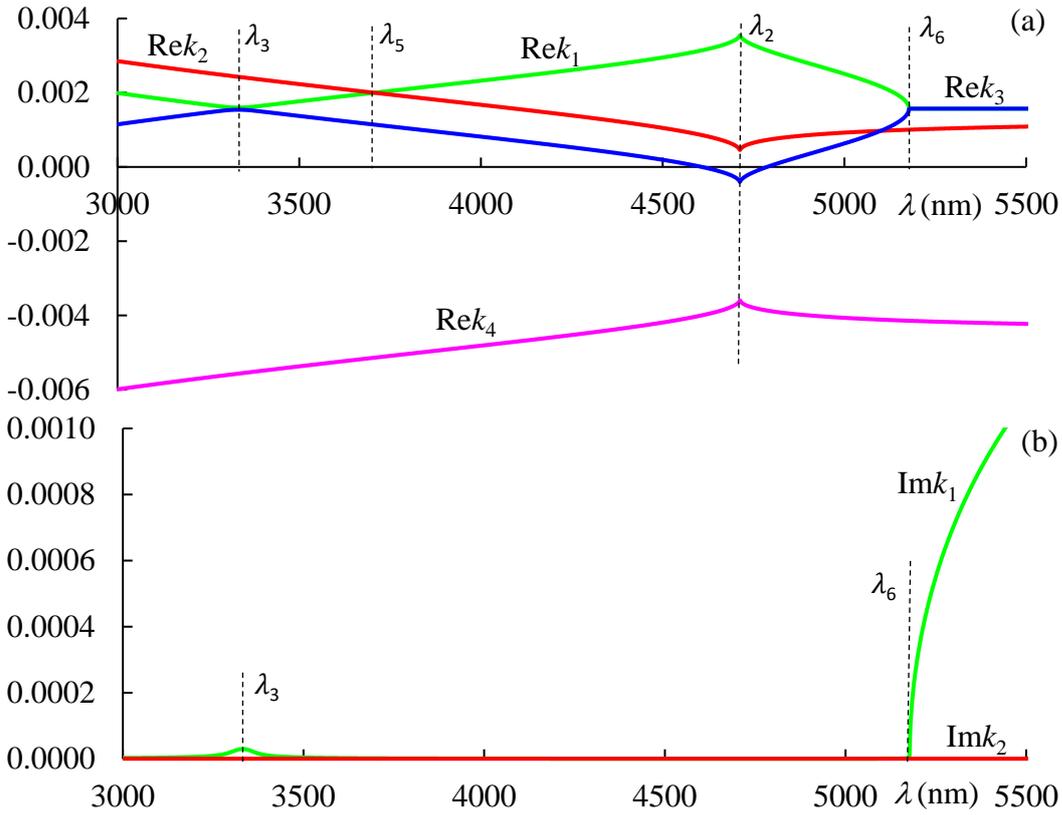

Fig. 5. The dependences of (a) real and (b) imaginary parts of wave vectors $k_{iz}$ on wavelength $\lambda$. $p = 4000$ nm, $\mathrm{Im}\varepsilon_1 = 0.0001$, $\mathrm{Im}\varepsilon_2 = 0$. The other parameters are the same as in Fig.1.

Now we pass to investigate the evolution of spectra of reflection, transmission and absorption at the change of helix pitch $p$ and of $B_{ext}$, for the most complete representation of the features of these spectra changes when changing the helix pitch and external static magnetic field.

Fig. 6 shows the evolution of the spectra of (a,b) reflection $R$, (c,d) transmission $T$ and (e,f) absorption for various values at the change of the helix pitch. The incident light has polarization coinciding with the first EP (left column) and with the second EP (right column). The directions of the incident light and the external magnetic field coincide.

These evolutions demonstrate the independence of the wavelengths determining the boundaries of the second and third PBGs, as well as the transmission band near the wavelength $\lambda_2$, from the pitch of the helix. This is expressed in the appearance of horizontal bands of red or blue in the spectra of reflection, transmission, and absorption for incident light with polarization coinciding with the second EP. In Fig. 6 b these areas are highlighted by white arrows. Further, Fig. 6 a, c, e demonstrates the transformation of the region of the first PBG into a narrow window of MIT with an increase in the pitch of the helix. Figure 6 c, d, e and f demonstrates two more interesting effects (see also above), namely, at a helix pitch of about 3150 nm, a broadband region of MIT is formed for the mode with the first EP and a region of MIA for the mode with the second EP (they are highlighted by white arrows). In the first case, a coherent perfect transmission occurs in a significant broadband region, and in the second case, coherent perfect absorption occurs again in broadband region. In this region, ideal unidirectional transmission takes place, too. That is here this CLC layer can work as tunable ideal optical diode.

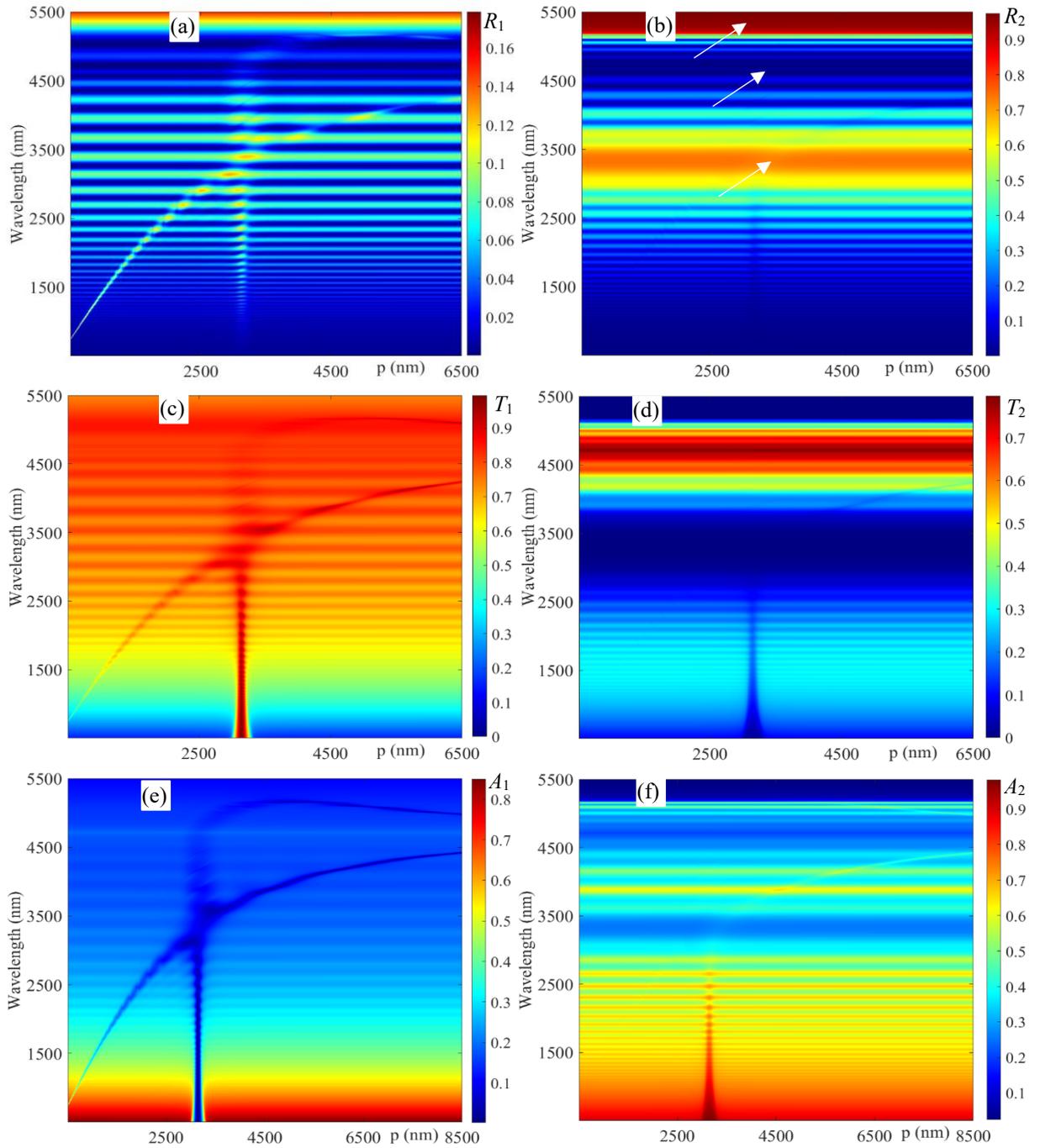

Fig. 6. The evolution of the spectra of (a,b) reflection $R$, (c,d) transmission $T$ and (e,f) absorption for various values at the change of the helix pitch. The incident light has polarization coinciding with the first EP (left column) and with the second EP (right column). The directions of the incident light and the external magnetic field coincide. $d=20p$. The other parameters are the same as in Fig.4.

And finally, Fig. 7 shows the evolution of the spectra of (a,b) reflection $R$, (c,d) transmission $T$ and (e,f) absorption at the change of external magnetic field . The incident light has polarization coinciding with the first EP (left column) and with the second EP (right column). Here $B = B_{ext} > 0$ means that the direction of light propagation and the direction of the external magnetic coincide, while $B_{ext} < 0$ means that these directions are reverse.

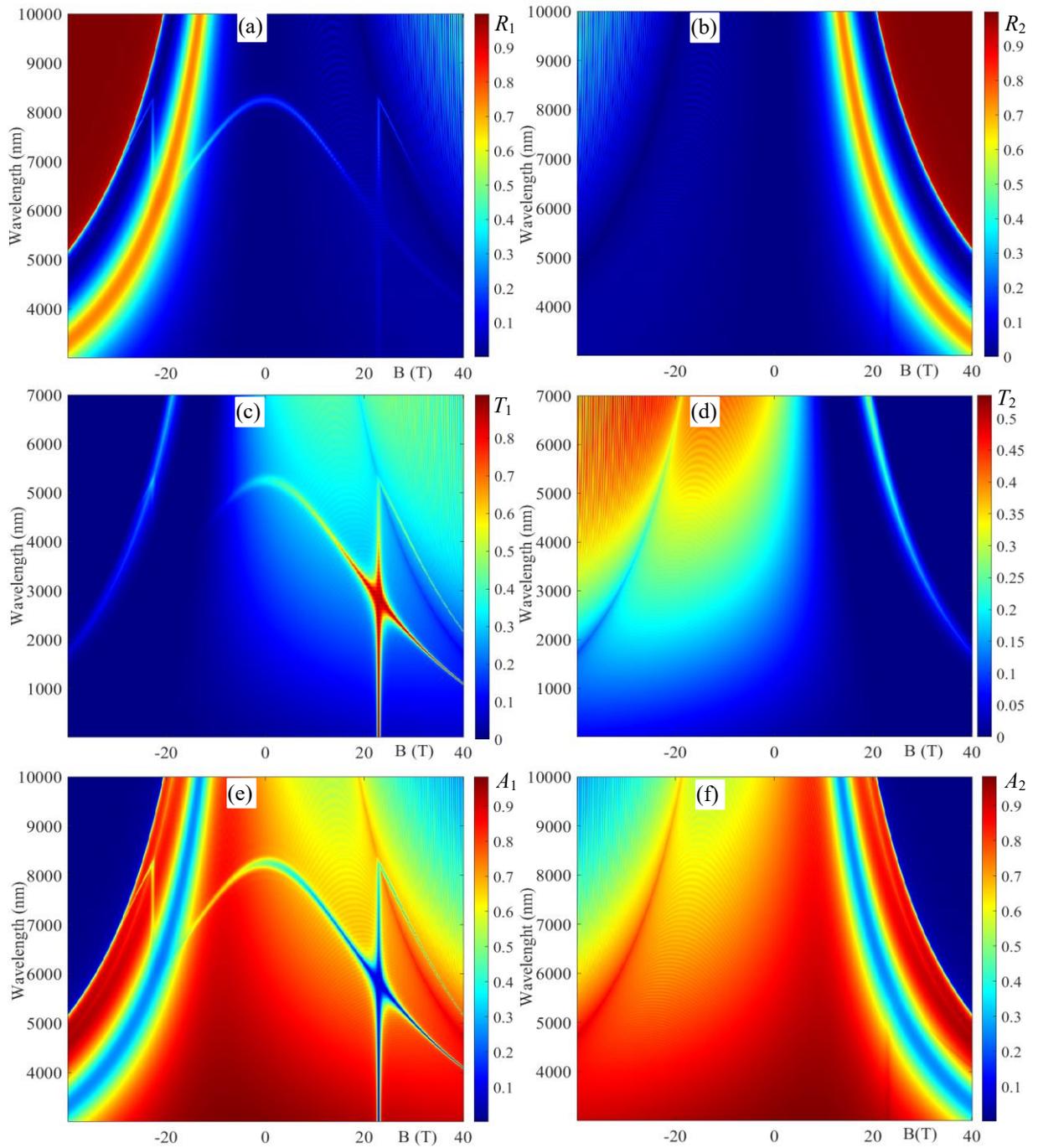

Fig. 7. The evolution of the spectra of (a,b) reflection $R$, (c,d) transmission $T$ and (e,f) absorption at the change of external magnetic field. The incident light has polarization coinciding with the first EP (left column) and with the second EP (right column). $p$=5500 nm. The other parameters are the same as in Fig.4.

### III. CONCLUSIONS

In conclusion, we investigated the properties of dichroic CLCs being in external static magnetic field directed along helix axis. We have shown that in the case of the wavelength dependence of magneto-optic activity parameter, new features appear in the optics of dichroic CLCs. First, we studied the behavior of the function g($\lambda$) and showed that it has two zeros and passes through a local maximum, and after the second zero it increases monotonically. We showed that in these points, that is on zeros and on local maximum of g($\lambda$) we have Dirac points, and moreover, in last Dirac point new type PBG arouse, which differs from the basic PBG (i.e., the PBG, that also exist in the absence of the external magnetic field). If the basic PBG is determined by the structure of the CLC, i.e., the chirality sign of the polarization of the incident diffracting light is determined only by the chirality sign of the CLC helix, then for the new one it is determined by the external magnetic field direction (i.e., on whether the directions of the external magnetic field and the incident light are parallel, or they are antiparallel). Near the second zero of g($\lambda$) a transmission band is formed for incident light with a certain polarization. And as for the second PBG the chirality sign of the polarization of the incident light for which this transmission band were formed is determined by the external magnetic field direction.

This system is non-reciprocal and the nonreciprocity changes over a wide range, it is observed both for reflection and transmittance and for absorption. The transformation of the region of the first (basic) PBG into a narrow window of MIT takes place with an increase in the pitch of the helix. With an increase of a helix pitch at a certain value of the latter, instead of a narrow line of MIT, a broadband region of MIT is formed. With a further increase in the helix pitch a new Dirac point appears and new MIT line appears. Then at $g \geq \varepsilon$ two from four wave vectors become complex at the absence of absorption and a new (the third) PBG will appears.

Finally, the soft matter nature of CLCs and their response to external influences lead to easily tunable multifunctional devices that can find a variety of applications. They can apply as tunable narrow band or broad band filters and mirrors, a highly tunable wideband/narrow-band coherent perfect absorber, transmitter, ideal optical diodes, and other devices The results obtained can find application in the design of innovative polarized optoelectronic devices, in particular, as devices for chemical analysis, biomedical diagnostics, polarization selectors, polarization imaging, and spectroscopic measurements.

**ACKNOWLEDGMENT**

This work was supported by the Foundation for the Advancement of Theoretical Physics and Mathematics "BASIS" (Grant No. 21-1-1-6-1).